# Shear and delamination behaviour of basal planes in $Zr_3AlC_2$ MAX phase studied by micromechanical testing

Siyang Wang[1,*], Oriol Gavalda-Diaz[2,1], Jack Lyons[1], Finn Giuliani[1]

[1] Department of Materials, Royal School of Mines, Imperial College London, London, SW7 2AZ, UK

[2] Composites Research Group, University of Nottingham, Nottingham, NG8 1BB, UK

*Corresponding author: siyang.wang15@imperial.ac.uk

## Abstract

The mechanical properties of layered, hexagonal-structured MAX phases often show the combined merits of metals and ceramics, making them promising material candidates for safety critical applications. While their unique mechanical performance largely arises from the crystal structure, the effect of chemistry on the properties of these materials remains unclear. To study this, here we employed two *in situ* electron microscope small-scale testing approaches to examine the micromechanical properties of $Zr_3AlC_2$, and compared the results with the properties of $Ti_3SiC_2$: we used micropillar compression tests to measure basal slip strength, and double cantilever beam splitting tests to evaluate fracture energy for basal plane delamination. We observed distinct and systematic differences in these measured properties between $Zr_3AlC_2$ and $Ti_3SiC_2$, where $Zr_3AlC_2$ appeared to be stronger but more brittle at the microscale, and discussed the implications of the results in the selection, design, and engineering of MAX phases for targeted engineering applications.

MAX phases are compounds with stoichiometry $M_{n+1}AX_n$ (n = 1/2/3), where M is an early-transition-metal element, A an A-group element, and X a C and/or N[1–3]. They have attracted interest during the past decades for their unique properties combining the merits of metals and ceramics[4,5]. Mechanically they exhibit high stiffness and strength, yet reasonable fracture toughness even at elevated temperatures[6], making them promising for high-temperature applications such as gas turbine and nuclear fuel cladding[7]. These properties are largely due to their layered, high c/a ratio, hexagonal structure, where $M_{n+1}X_n$ and A layers in the basal planes stack on top of each other in alternating manner. Although structurally anisotropic, the crystals have multiple ways to accommodate external load, where the basal planes play a central role: they can slip, kink, and delaminate depending on the stress state[8–10]. The ductility of MAX phases at the macroscale is mainly achieved through basal slip, and toughness is largely dominated by basal plane delamination. Therefore the mechanical performance of these materials is fundamentally determined by the microscale behaviour of basal planes.

To understand the micromechanical properties of a certain deformation/fracture mode, it is essential to test single crystals with appropriate orientations under well-defined stress states[11]. This is particularly useful for materials like MAX phases where making pure bulk samples is difficult, therefore results of macroscopic tests is contributed also by the secondary phases like binary carbides, leading to uncertainties in data analysis. For plastic deformation this is often achieved through micropillar compression[12–14] or microcantilever bending[15] tests, while for fracture, single [16] or double[4,17] cantilever beams (DCB) are commonly used. Previously, we, and a few others, have demonstrated that micromechanical testing is a powerful tool for extracting the fundamental properties of 312 MAX phases (312 stands for the $M_3AX_2$ stoichiometry) such as $Ti_3SiC_2$, including critical resolved shear stress (CRSS) for basal slip[5] and fracture energy for basal plane delamination[4]. What has not been studied is the effect of chemistry on these properties. For example, the relatively novel $Zr_3AlC_2$ MAX phase[1] has the same crystal structure as $Ti_3SiC_2$, but how important the composition is to the micromechanical properties remains unknown. This triggered the motivation of this work where we evaluated the micromechanical properties of $Zr_3AlC_2$, and compared them with those of $Ti_3SiC_2$.

$Zr_3AlC_2$ was synthesised through hot pressing mixed, ball milled, and pre-compacted powders of $ZrH_2$, Al and C (in graphite form), in a FCT Systeme GmbH vacuum hot press using parameters modified from those reported in Ref.[1]. We also synthesised and tested $Ti_3SiC_2$ for comparison with $Zr_3AlC_2$. Details about the synthesis of $Ti_3SiC_2$ can be found in prior work[4], which reported a method analogous to that in Ref.[18]. The samples were mechanically ground with SiC abrasive papers, polished with diamond paste, and finally vibratory polished with Buehler MasterPolish suspension for characterisation with electron microscopy.

Electron backscatter diffraction (EBSD) was used to extract the crystal orientations of the grains in the samples, in order to pick out grains with desired orientations for micromechanical testing. For micropillar compression tests, the grains of interest are those with their basal planes oriented at ~45° to the sample surface, for activating basal slip during uniaxial compression. For DCB splitting tests, we chose the grains with their basal planes perpendicular to the sample surface, and made the samples such that the notches run along the basal plane. EBSD scanning was performed on a Thermo Fisher Scientific (TFS) Quanta 650 scanning electron microscope (SEM) equipped with a Bruker eFlashHR (v2) EBSD camera, using a beam acceleration voltage of 20 kV and a probe current of ~10 nA.

Fabrication of small-scale specimens for micromechanical testing was achieved through focussed ion beam (FIB) milling on a TFS Helios 5 CX DualBeam microscope, using a FIB acceleration voltage of 30 kV. Beam currents from ~7 nA to ~200 pA were employed for the milling of micropillars and DCBs. Higher beam currents were used to create trenches around samples, thereby enabling real-time visualisation of subsequent *in situ* testing processes. Lower beam currents (~500 to 200 pA) were used for final tailoring the contour of the samples. The taper angle of the pillars was measured to be ~4.5°. For the DCBs, notches were cut on the top surface of the samples using a line scan at 10 pA beam current.

*In situ* SEM micromechanical testing was performed using a displacement-controlled Alemnis nanoindenter on the TFS Quanta SEM. The micropillars were compressed with a circular flat punch indenter to achieve a (near) uniaxial stress state, while the DCBs were split using a 60° wedge indenter to drive stable crack growth.

For the DCB tests, the critical energy release rate in Mode I fracture of the basal plane for the $Zr_3AlC_2$ single crystals was extracted using the following equation as per prior work[4,19]:

$$G_I^c = \frac{3E_{11}\delta^2 w^3}{4(a+\chi w)^4}\left[1 + (1+\nu_{13})\left(\frac{w}{a+\chi w}\right)^2\right] \qquad 1$$

where *a* is the crack length, and *δ* and *w* are the horizontal displacement and width of the beams, respectively. These values were measured from the real-time SEM images recorded during the tests. *χ* is a correction factor accounting for rotation at the crack tip, and for the material studied it is defined based on the elastic constants as:

$$\chi = \sqrt{\frac{a_{66}}{11a_{11}}}\sqrt{3 - 2\left(\frac{\Gamma}{1+\Gamma}\right)^2}, \quad with\ \Gamma = 1.18\frac{a_{66}}{\sqrt{a_{11}a_{22}}} \qquad 2$$

where $a_{11} = 1/E_{11}$, $a_{22} = 1/E_{33}$ and $a_{66} = 1/G_{13}$, and $E_{ii}$ and $G_{ii}$ are the elastic constants for $Zr_3AlC_2$, calculated from the components of the compliance matrix reported in Ref.[20], which were obtained through first principles calculation.

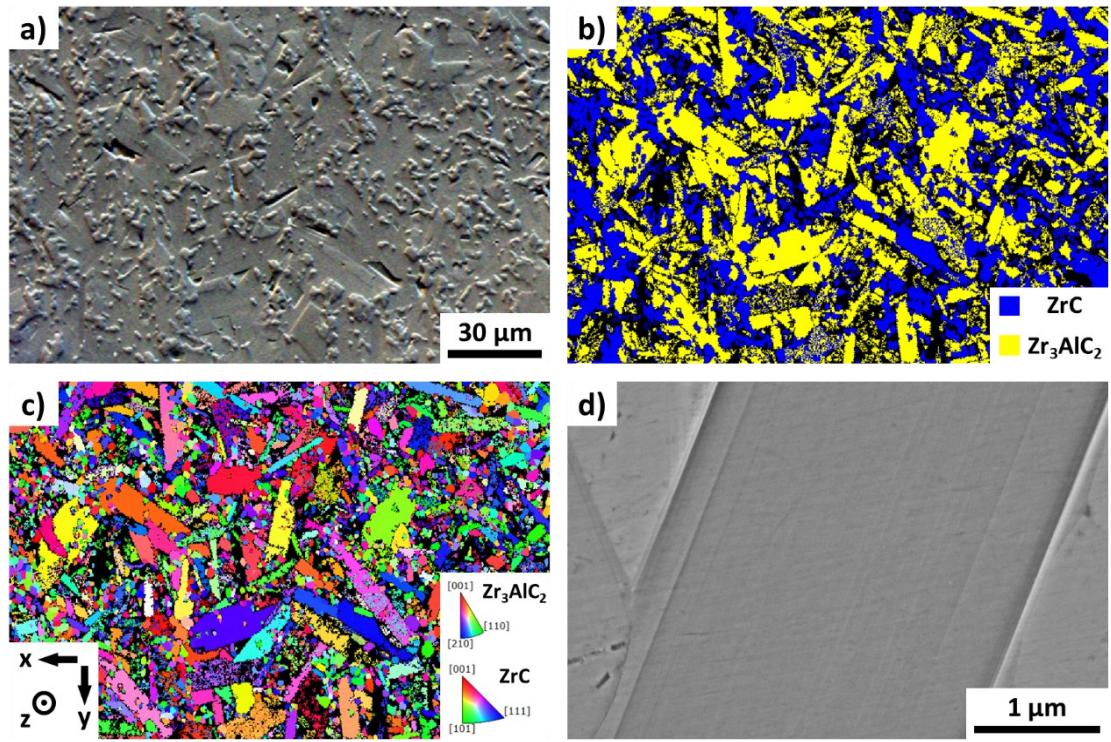

Figure 1 Microstructure of the $Zr_3AlC_2$ MAX phase synthesised and studied in this work. (a) Argus™ forescatter electron image, (b) EBSD phase map and (c) EBSD IPF-*z* map of the material showing $Zr_3AlC_2$-ZrC dual phase microstructure. (d) Backscattered electron image showing the structure of a $Zr_3AlC_2$ lath.

The results of the SEM/EBSD characterisation of the synthesised material are shown in Figure 1(a-d), where $Zr_3AlC_2$ laths can be observed in the $Zr_3AlC_2$-ZrC dual phase microstructure. Small-scale single crystal micropillars fabricated in those grains showing high Schmid factors (>0.45) for basal slip were then tested *in situ*, and the post-deformation SEM images of the pillars as well as the mechanical responses are shown in Figure 2. All the pillars tested, including both $Zr_3AlC_2$ and $Ti_3SiC_2$, exhibited basal slip. The slip directions of the pillars, as determined using the post-deformation SEM images, were all along the $<a>$ (or $<11\bar{2}0>$) crystallographic direction of each sample, confirming the activation of $<a>$ basal slip. The vertical positions of the slip bands are not all at the very top of the pillars, indicating that the taper did not massively influence the deformation behaviour. There are, however, distinct differences in the mechanical responses between the two MAX phases:

1. The "yield stresses", defined as the stresses at the end points of linear elastic regions, are systematically higher for $Zr_3AlC_2$ (2-4 GPa) than for $Ti_3SiC_2$ (<1 GPa).
2. All the $Zr_3AlC_2$ pillars exhibited significant stress drops shortly after yield, whereas the $Ti_3SiC_2$ pillars showed only few subtle stress drops and much more stable plastic flow upon further loading.

3. For the $Zr_3AlC_2$ pillars, the flow stress levels after the stress drops did not get back to the levels prior to the drops, and this is different to the flow behaviour of the $Ti_3SiC_2$ pillars where in most cases the stress levels recovered after dropping.

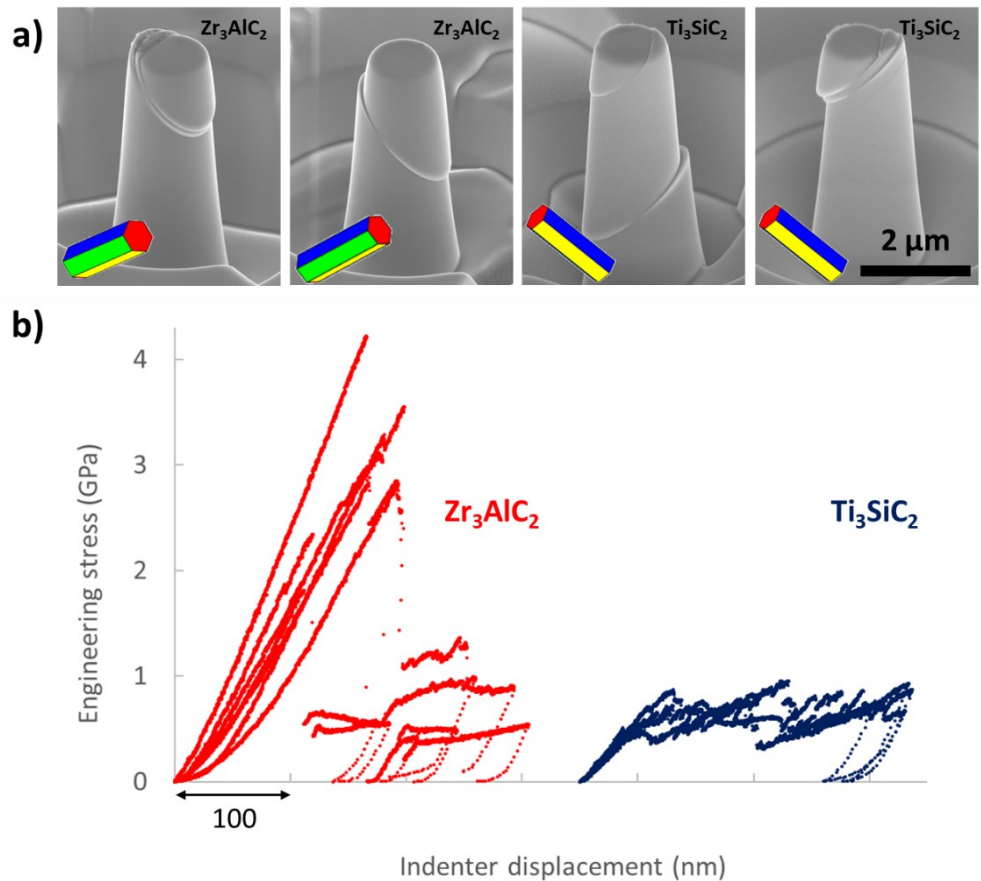

Figure 2 Results of the micropillar compression tests. (a) Post-deformation SEM images of some of the $Zr_3AlC_2$ and $Ti_3SiC_2$ micropillars tested showing basal slip, with inserts of unit cell representations of crystal orientations, oriented with respect to the viewing angle. (b) Mechanical responses (stress-displacement curves) of the $Zr_3AlC_2$ (red) and $Ti_3SiC_2$ (blue) micropillars recorded during the tests, showing distinct differences in yield stress and plastic flow behaviour. These overlaid curves are also plotted separately in Figure S2.

The CRSS for basal slip at this length scale were calculated from the yield stresses and the crystal orientations (hence Schmid factors) of the pillars, and the results are plotted in Figure 3. Our better ability in routinely growing large grains in $Ti_3SiC_2$ than $Zr_3AlC_2$ allowed us to also study the size dependence of the CRSS for basal slip in $Ti_3SiC_2$, through testing pillars with mid-height diameters of 1, 1.7 and 5 µm (post-deformation SEM images of some of the pillars are given in Figure S1). Through least-squares fitting of the data in log-log scale, the relationship between the CRSS for basal slip and the pillar size (mid-height diameter) can be described as

$$\tau_{CRSS, basal\ slip, Ti_3SiC_2} = 640 d^{-1.43} \qquad\qquad 3$$

which showed good agreement with the results reported in Ref.[5] (the orange line in Figure 3) where the size effect likely arose from defect population within the pillars (that smaller test pieces contained less pre-existing defects than larger ones)[21]. It is well-known that results of small-scale micropillar compression tests can be sensitive to experimental factors such as sample geometry, exact stress state (or tip-sample alignment), and defect population within materials. The good agreement between our results and those of Higashi et al.[5] across length scales therefore gives confidence for the validity of both work, and suggests that our observations (the dramatic yield stress difference in particular) are unlikely due to testing artefacts. Comparing the data for $Zr_3AlC_2$ with that for another material also allowed deformation behaviour to be contrasted as shown above.

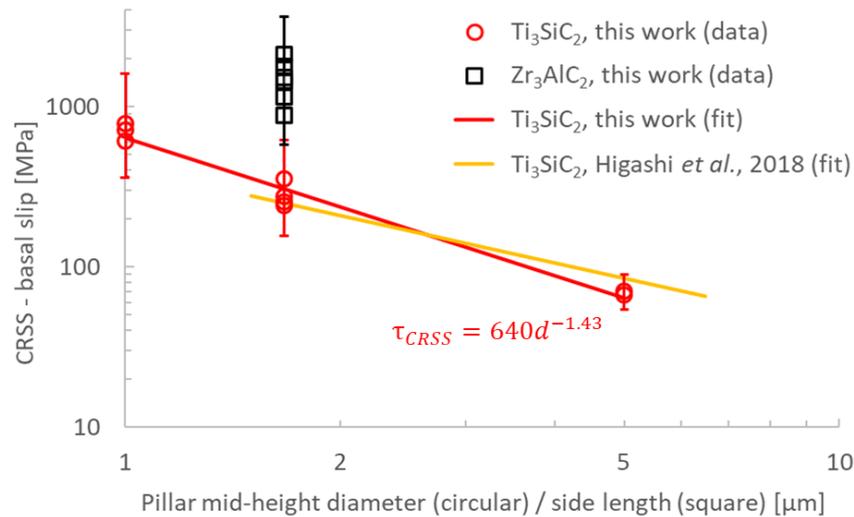

Figure 3  Comparison of the CRSS for basal slip between $Zr_3AlC_2$ and $Ti_3SiC_2$, and the size dependence of the CRSS for basal slip in $Ti_3SiC_2$, measured in the present work and by Higashi *et al.*[5]. Error bars represent uncertainties in CRSS determination associated with the taper of the micropillars.

The CRSS for basal slip in $Zr_3AlC_2$ is over 3 times that in $Ti_3SiC_2$, when comparing the results obtained from the 1.7 µm-diameter pillars for both materials (Figure 3). This means that at the microscale, basal slip in $Zr_3AlC_2$ is stronger than in $Ti_3SiC_2$, which implies the potential for $Zr_3AlC_2$ to show a higher level of mechanical strength than $Ti_3SiC_2$ at the large scale if the defect population is carefully controlled through microstructural engineering. We note that the deformation behaviour (in terms of the slip band structure post-deformation, and the flow stress level post-yield) of the $Zr_3AlC_2$ pillars does not differ dramatically from that of the $Ti_3SiC_2$ pillars. However, the vastly significant difference in CRSS for basal slip between the two phases would suggest that, at this length scale, basal planes in $Ti_3SiC_2$ will not sustain a shear stress of, say, 800 MPa that may not even cause $Zr_3AlC_2$ to plastically deform. This highlights the contrasting abilities of the two phases in accommodating external load.

Upon intrinsically displacement-controlled micropillar compression tests, significant stress drops without recovery of the stress level, could happen in the following scenarios:

- The sample plastically deformed through deformation twinning[22]. However, for the pillars in Figure 2(a), the clear slip bands along the layers rule out this possibility.
- In metallic samples when dislocations are pinned by solute atoms and therefore require an extra stress for source activation. This might be possible for $Zr_3AlC_2$, if local chemical inhomogeneities are present causing lattice distortions, which could retard dislocation motion. To our best knowledge, however, there is no experimental evidence in the literature indicating the presence or absence of such compositional variations in $Zr_3AlC_2$, and this could be an interesting topic for future studies through analysis with atom probe tomography.
- When dislocations have to fracture a surface layer before they can traverse it, such as an air- or FIB-induced coating on the pillar surface. Although the thermodynamic stability of $Zr_3AlC_2$ was shown in the literature[23,24], $Al_2O_3$ scales can form on the surface of certain Al-containing MAX phases such as $Ti_2AlC$, $Ti_3AlC_2$ and $Cr_2AlC$ upon air exposure at elevated temperatures[7]. At room temperature, the thickness of the oxide scale formed on Al metal saturates at ~5 nm[25]. The room-temperature hardness of the $Al_2O_3$ layer formed on the surface of $Cr_2AlC$ (after oxidation at 1200 °C for over 29 h) is almost equal to that of the bulk[26]. FIB-induced surface damage layer may also be present, and the thickness is typically on the order of 10 nm for 30 kV $Ga^+$ FIB[27]. However, this effect of the FIB-induced damage layer on the mechanical behaviour was typically observed on micro/nanopillars of metals/alloys[28,29] whose intrinsic strength values are an order of magnitude lower than the ceramic material studied here. Thus we doubt if such layers of approximately 10 nm (on $Zr_3AlC_2$ pillars of 1.7 µm diameter) would be able to cause the GPa-level stress drops observed in Figure 2. Future TEM work would be helpful to clarify if these layers exist, and if so, their contributions to the observed yield behaviour of $Zr_3AlC_2$.
- The sample experienced quasi-brittle failure. In our prior work on an intermetallic compound[14], we observed highly similar mechanical responses of single crystal micropillars (showing significant stress drops without full stress recovery). TEM imaging revealed both dislocations and microcracks along the slip band. Hence the stress drops could happen upon the formation of microcracks along the slip bands, potentially due to local high stresses by virtue of dislocation pile-up. This is supported by the fact that one out of the seven $Zr_3AlC_2$ pillars tested shear fractured during the test (post-test SEM image and stress-displacement curve shown in Figure S3 and Figure S2(g), respectively). Meanwhile none of the $Ti_3SiC_2$ pillars was found to fracture during the tests. Nonetheless confirming the detailed mechanism unambiguously will be a focus of future work using techniques such as in situ mechanical testing in the TEM, which would be able to present evidence of dislocation activity and microcracking in the slip planes, and also to examine if dislocation obstacles and/or a surface layer are present.

Conversely, when the mechanical response shows that stress levels recovered after the stress drops (the results for the $Ti_3SiC_2$ pillars here for example), it is more likely that the sample experienced pure plastic deformation *via* slip. Stress drops (or strain bursts in load-controlled tests) are frequently observed during slip of micropillars[30]. In those cases, however, the overall strengths of the samples are usually retained. This is due to the nature of plasticity where no significant reduction in the total number of bonds in the material occurred at low levels of strains. Therefore, the measured stress can often get back to the original level when the sample is reloaded upon further test[13].

Therefore, the difference in the extent of stress drops between the two phases motivated us to investigate the potential differences in fracture properties. The mode I fracture behaviour and toughness of basal planes in $Zr_3AlC_2$ were studied with DCB splitting tests shown in Figure 4, where a crack can be observed to grow stably along the basal plane of $Zr_3AlC_2$ without deflection up until ~4 μm. Three tests all showed stable and straight crack growth which allowed extraction of the fracture energy plotted in Figure 5.

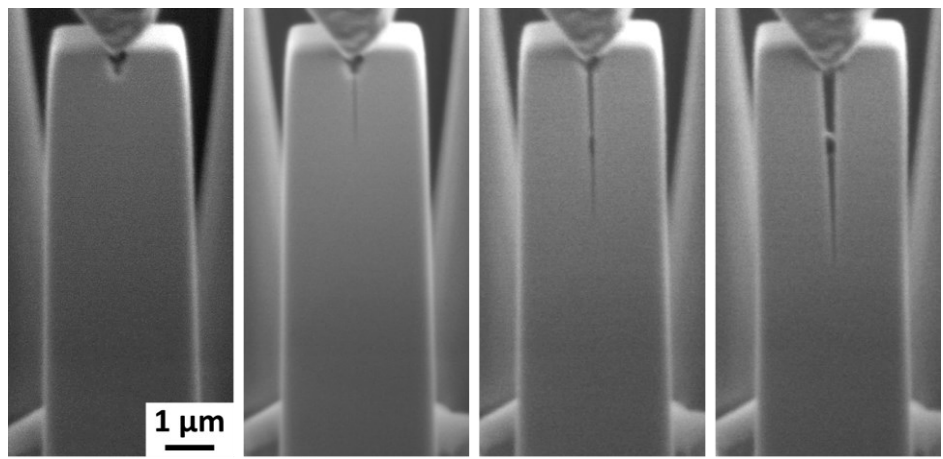

Figure 4  Real-time SEM images recorded during the DCB splitting tests showing stable crack growth along the basal plane in $Zr_3AlC_2$.

Broadly, the three tests yielded similar fracture energy values of 5.5 ± 2.5 $J/m^2$. This fracture energy range is evidently lower than that measured for $Ti_3SiC_2$ (10.0 ± 2.0 $J/m^2$, the shaded area in Figure 5) in our prior work using the same method[4]. Our results therefore indicate that although the 312 MAX phases share the hexagonal crystal structure which fundamentally leads to their unique mechanical properties, the chemistry of the materials also has a pronounced effect on the microscale strength and toughness: for instance, the CRSS for basal slip of $Zr_3AlC_2$ at the micron scale is over 3 times that of $Ti_3SiC_2$, but the energy for mode I fracture on the basal plane for $Zr_3AlC_2$ is only about half that for $Ti_3SiC_2$. This suggests that MAX phases can potentially provide a reasonably broad spectrum of mechanical properties, and therefore the various materials of this class may be suitable or customised for a wide range of applications requiring specific properties. More importantly, this also implies that the composition and microstructure of these materials could possibly be engineered to optimise the combination of strength and toughness, thereby making new materials with desired properties and potentially breaking through the strength-ductility trade-off of engineering materials.

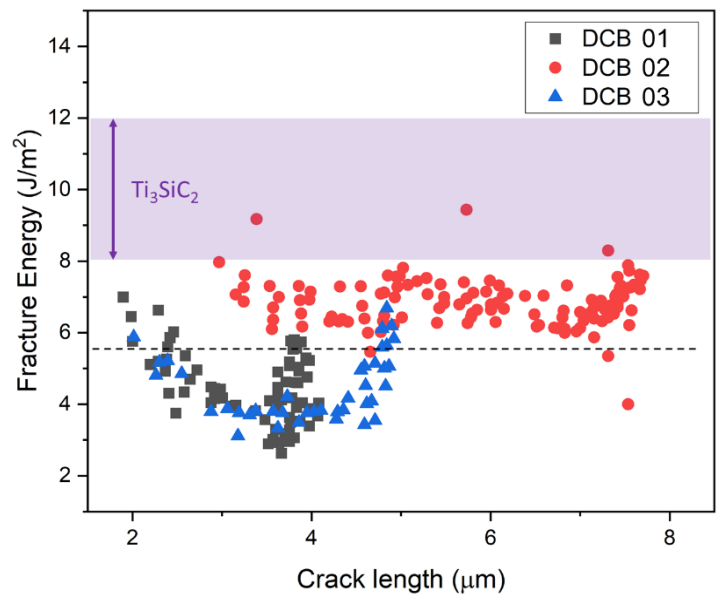

Figure 5  Energy for mode I fracture on the basal plane in $Zr_3AlC_2$ over crack length, obtained for three different $Zr_3AlC_2$ DCBs, all oriented with their basal planes parallel to the notches. The purple-coloured shaded area on the plot corresponds to the basal plane fracture energy range measured for $Ti_3SiC_2$ in prior work[4].

## Acknowledgements

SW, OGD and FG acknowledge funding from Engineering and Physical Science Research Council (EPSRC) through MAPP project (EP/P006566/1). JL and FG acknowledge funding from the EPSRC Centre for Doctoral Training in Nuclear Energy: Building UK Nuclear Skills for Global Markets (EP/L015900/1). The TFS Quanta SEM used was supported by the Shell AIMS UTC and is housed in the Harvey Flower EM suite at Imperial College London. The TFS Helios microscope used is part of the cryo-EPS facility at Imperial College London funded by EPSRC (EP/V007661/1).

# Supplementary figures

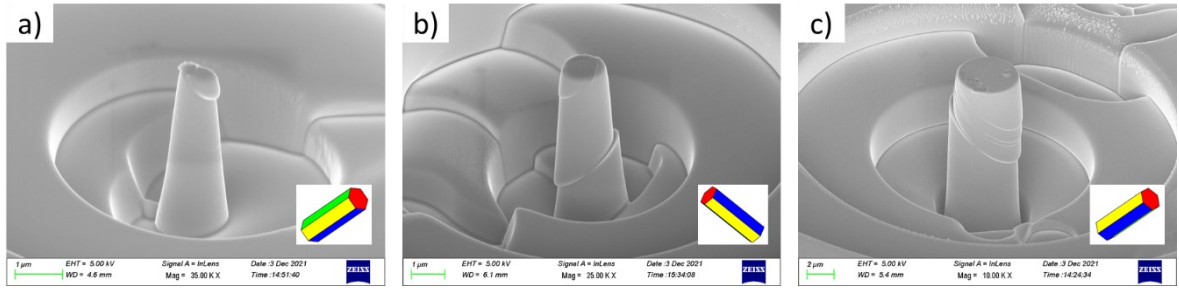

Figure S1  Post-deformation SEM images of (a) 1 µm, (b) 1.7 µm and (c) 5 µm mid-height diameter $Ti_3SiC_2$ micropillars tested showing basal slip, with inserts of unit cell representations of crystal orientations, oriented with respect to the viewing angle.

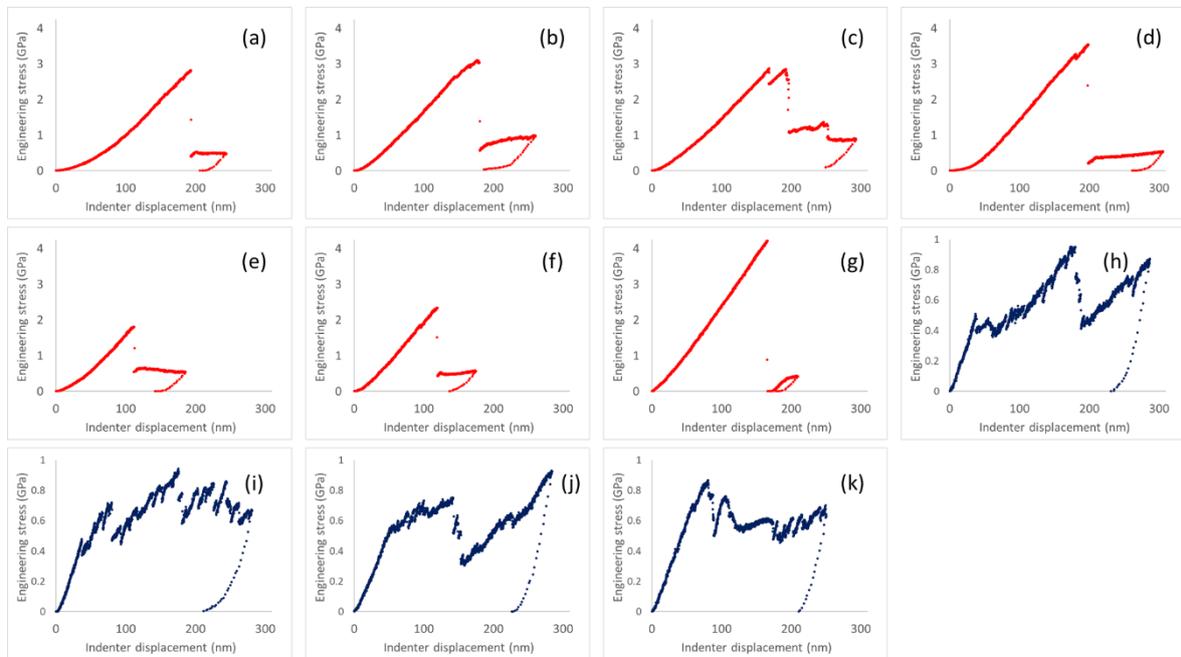

Figure S2  Engineering stress – indenter displacement curves recorded during compression tests of (a-g) $Zr_3AlC_2$ and (h-k) $Ti_3SiC_2$ micropillars with 1.7 µm mid-height diameter.

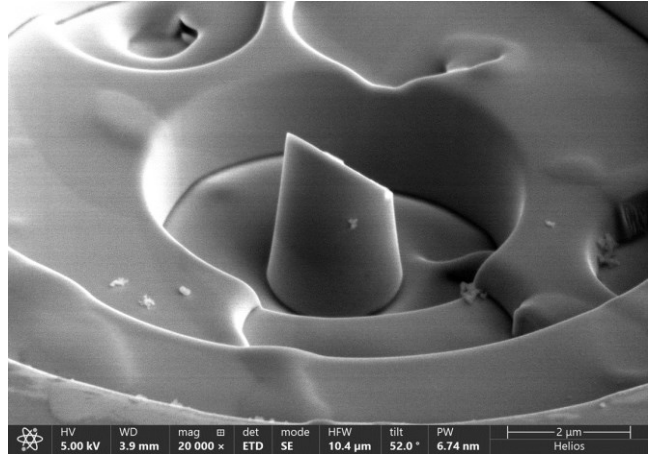

Figure S3 Post-test SEM image of a Zr$_3$AlC$_2$ pillar that shear fractured along the basal plane during the test. The corresponding stress-displacement curve is shown in Figure S2(g).